\documentclass[12pt]{article}

\usepackage[T1]{fontenc}
\usepackage[utf8]{inputenc}
\usepackage{authblk}

\usepackage{graphicx}
\usepackage{slashed}
\usepackage{feynmp}
\usepackage{multirow}
\usepackage{epstopdf}
\usepackage[body={17.5cm, 21cm},right=2cm]{geometry}
\usepackage{amssymb}
\usepackage{amsmath}
\usepackage{subcaption}
\usepackage{verbatim}
\usepackage{authblk}
\usepackage{booktabs}
\usepackage{fancyhdr}
\usepackage{ctable}
\usepackage{cancel}
\usepackage[boxsize=2em,aligntableaux=center]{ytableau}
\usepackage{dsfont}

\numberwithin{equation}{section}

\newcommand\eea{\end{eqnarray}}
\newcommand\bea{\begin{eqnarray}}

\def\beq{\begin{equation}}
\def\eeq{\end{equation}}

\newcommand{\be}{\begin{equation}}
\newcommand{\ee}{\end{equation}}
\newcommand{\ba}{\begin{align}}
\newcommand{\ea}{\end{align}}
\newcommand{\bg}{\begin{gather}}
\newcommand{\eg}{\end{gather}}
\newcommand{\bseq}{\begin{subequations}}
\newcommand{\eseq}{\end{subequations}}

\newcommand{\Tr}{{\rm Tr}}

\newcommand{\yss}{\ytableausetup{boxsize=0.5em}}
\newcommand{\ysn}{\ytableausetup{boxsize=2em}}
\newcommand{\symm}{\yss\ydiagram{2}\ysn}
\newcommand{\asymm}{\yss\ydiagram{1,1}\ysn}
\newcommand{\fund}{\yss\ydiagram{1}\ysn}
\newcommand{\adj}{\ensuremath{\text{Adj}}}

\begin{document}

\begin{titlepage}

\title{\LARGE\bfseries\bf New self dualities and duality cascades}
\author{Anson Hook}
\affil{\small\slshape School of Natural Sciences\\
 Institute for Advanced Study\\
Princeton, NJ 08540}
\maketitle

\begin{abstract}

New self-dualities involving two index tensors are derived.  These new self-dualities are used to build various duality cascades.  Both vector like and chiral cascades are presented.  Aside from ending in confinement, these duality cascades can also end in interacting conformal field theories, free field theories, and meta-stable supersymmetry breaking.  Higgsing effects are built into the self-duality so that when the gauge groups are small enough, supersymmetry is broken through the rank condition.  Dynamical supersymmetry restoration occurs far from the SUSY breaking vacuum resulting in a long lived meta-stable vacuum.  It is found that Coulomb branches are 
critical in the stabilization of runaways and dynamical supersymmetry restoration. 

\end{abstract}

\vspace{1cm}

\end{titlepage}

\tableofcontents

\section{Introduction}

The Klebanov-Strassler(KS) cascade\cite{Klebanov:2000hb,Klebanov:2000nc,Strassler:2005qs} has been fruitful ground for better understanding ADS/CFT\cite{Aharony:1999ti}.  The gauge theory description has a $SU(N) \times SU(N+M)$ gauge theory undergoing a duality cascade.  As the gauge groups become strongly coupled, Seiberg duality\cite{Seiberg:1994pq} is applied resulting in an identical theory but with smaller gauge groups.  The UV has ever increasing numbers of degrees of freedom, while the IR is a gapped theory.  The gravity description is a deformed conifold where the singularity is screened, geometerizing confinement. 

Cascades are useful for supersymmetry(SUSY) breaking scenarios.  
Cascades involving SUSY breaking can be used for model building in extra dimensions\cite{Randall:1999ee} or for building nonsupersymmetric string vacua\cite{Kachru:2003aw}.  Many attempts have been made to obtain cascades which break supersymmetry.  Some of these involve deforming the KS cascade by adding anti-D3 branes\cite{Kachru:2002gs} yielding a meta stable SUSY breaking vacuum.  Others use different backgrounds finding SUSY breaking in the form of runaways\cite{Franco:2005zu,Berenstein:2005xa,Bertolini:2005di,Brini:2006ej,Herzog:2004tr} which in some cases can be stabilized by hand at the end of the cascade\cite{Argurio:2006ew}.  Both of these attempts have their drawbacks, whether it be runaway behavior rather than SUSY breaking\cite{Intriligator:2005aw}, singular geometries\cite{Dymarsky:2011pm,Bena:2009xk}, the lack of a field theory description, or the lack of a controllable gravity dual.

Rather than deforming a cascade to break supersymmetry, this paper takes a different approach.  We build the SUSY breaking mechanism into the RG cascade itself in the form of Higgsing effects.  Higgsing is introduced into a  known duality by adding a term linear in the meson.  In the IR, the superpotential becomes
\bea
W = \Tr M + q M \tilde q
\eea
where $q$ and $\tilde q$ are the dual quarks with a gauge group of size $\tilde N_c$ and $N_f$ flavors.  The rank of $q \tilde q$ is determined by min($\tilde N_c$,$N_f$).  If $\tilde N_c > N_f$ the quarks obtain a vacuum expectation value(vev) that Higgses the gauge group reducing its from $\tilde N_c$ to $\tilde N_c - N_f$.
If $\tilde N_c < N_f$, it is impossible to set $q \tilde q \sim \mathds{1}_{N_f \times N_f}$ and supersymmetry is broken by the rank condition\cite{Intriligator:2006dd}.  As in the original ISS situation, supersymmetry is dynamically restored by gaugino condensation at large field values.

  The rank condition can be built into a cascade if self-dualities are found where Higgsing effects are crucial.  In the Klebanov-Strassler cascade, the cascade proceeds as $SU(N) \times SU(N-M) \rightarrow SU(N-2M) \times SU(N-M) \rightarrow SU(N-2M) \times SU(N-3M) \rightarrow \cdots$ until one of the gauge group goes negative.  A negative gauge group indicates that Seiberg duality has been incorrectly applied and that instead there is an ADS-superpotential.  However, if Higgsing is important to each step of these dualities, then the gauge group becoming negative is instead an indication that the rank condition is no longer satisfied so that a meta-stable supersymmetric vacuum is present.
  
In the large N limit, the gauge group is always large enough so that the quarks obtain a supersymmetric vev.  At the bottom of the cascade, the gauge groups are small so that the rank condition is not satisfied and supersymmetry is broken.  This approach gives a dynamical reason why the SUSY breaking vacuum is present in the IR.  This paper constructs several cascades which exhibit this feature.

Another interesting feature found is the importance of branches of moduli space where the gauge symmetry is broken to subgroups.  A runaway is found which is not stabilized in the magnetic theory, but is stabilized in the electric theory on other branches with smaller gauge groups.  A similar effect is found in dynamical supersymmetry restoration, where the supersymmetric vacuum is caused by gaugino condensation of a subgroup of the entire gauge symmetry.

  The paper is organized as follows.  Sec.~\ref{Sec: dualities} discusses several new gauge theory self-dualities where Higgsing is important with a more exhaustive list presented in App.~\ref{App: SOSP}.  Sec.~\ref{Sec: adjoint cascade} presents a cascade involving adjoints which behaves very similarly to the Klebanov-Strassler cascade.  Sec.~\ref{Sec: chiral cascade} presents a chiral cascade where the IR features meta-stable supersymmetry breaking.  Sec.~\ref{Sec: free cascade} presents a cascade where the self-dual point is a free field theory.  Finally Sec.~\ref{Sec: conclusion} concludes with future directions.

\section{Self-dual theories with two index tensors}
\label{Sec: dualities}

In this section, we present several self-dualities where the relevant gauge theories have two index tensors.  The approach used to derive these new self-dualities is Higgsing effects.  As a result, when these self-dualities are incorporated into a duality cascade, the Higgsing effects generate meta-stable supersymmetry breaking through the rank condition.
  This section presents self-dualities for SU gauge groups with either an adjoint or both a symmetric and antisymmetric tensor.  Additional self-dualities are presented in App.~\ref{App: SOSP}.

\subsection{$SU(N_c)$ with a single adjoint}
\label{Sec: SU adj}

A duality for SU gauge groups with an adjoint and $N_f$ flavors was found in Ref.~\cite{Kutasov:1995ve,Kutasov:1995np} and is summarized below.  The electric theory is 
\begin{center}
\be
\begin{tabular}{c|c|cccc}
&$SU(N_c)$&$SU(N_f)_L$&$SU(N_f)_R$&$U(1)_B$&$U(1)_R$\\
\hline
&&\\[-8pt]
$Q$ & $\fund$ & $\fund$ & & $1$ & $1-\frac{2 N_c}{(k+1) N_f}$ \\
$\tilde Q$ & $\overline \fund$ &  & $\overline \fund$ & $-1$ & $1-\frac{2 N_c}{(k+1) N_f}$ \\
$X$ & $\adj$ &  & &  & $\frac{2}{k+1}$ 
\end{tabular}
\ee
\end{center}
\bea
\label{Eq: leading}
W = \frac{\lambda}{k+1} \Tr X^{k+1}
\eea
This theory is dual to 
\begin{center}
\be
\begin{tabular}{c|c|cccc}
&$SU(\tilde N_c)$&$SU(N_f)_L$&$SU(N_f)_R$&$U(1)_B$&$U(1)_R$\\
\hline
&&\\[-8pt]
$q$ & $\fund$ & $\overline \fund$ &  & $\frac{N_c}{k N_f - N_c}$ & $1-\frac{2}{k+1} \frac{k N_f - N_c}{N_f}$\\
$\tilde q$ & $\overline \fund$ &  & $\fund$ & $-\frac{N_c}{k N_f - N_c}$ & $1-\frac{2}{k+1} \frac{k N_f - N_c}{N_f}$ \\
$x$ & $\adj$ &  & & & $\frac{2}{k+1}$ \\
$M_j = Q X^j \tilde Q$ & & $\fund$ & $\overline \fund$ & & $2-\frac{4 N_c}{(k+1) N_f} +\frac{2j}{k+1}$
\end{tabular}
\ee
\end{center}
\bea
\nonumber
W = - \frac{\lambda}{k+1} \Tr x^{k+1} +  \frac{\lambda}{\mu^2}\sum_{j=0}^{k-1} M_{k-j-1} q x^j \tilde q
\eea
where the index j can run from $0$ to $k-1$ and $\tilde N_c = k N_f - N_c$.  The auxiliary scale $\mu$ is present for dimensional reasons.  The theory has a runaway if $N_f < \frac{N_c}{k}$\cite{Kutasov:1995np}.  The argument for a runaway is to study the electric theory with the superpotential
\bea
\label{Eq: sub}
W = \sum_{i=0}^k \lambda_l \Tr X^{i+1}
\eea
Adding in the subleading terms does not affect the large vev behavior.  If there is a runaway for the superpotential shown in Eq.~\ref{Eq: sub}, then there is a runaway for the superpotential shown in Eq.~\ref{Eq: leading}.  The potential resulting from Eq.~\ref{Eq: sub} has $k$ solutions.  Consider giving an expectation value to $X$.  These expectation values are labeled by integers $i_l$ which label how many of its eigenvalues are in the $l$th solution and obey $\sum_{l=1}^k i_l = N_c$.  The gauge group is broken down to $SU(N_c) \rightarrow SU(i_1) \times SU(i_2) \times \cdots SU(i_k) \times U(1)^{k-1}$.  The adjoint is massive and integrated out while each gauge group has $N_f$ flavors.   An ADS superpotential\cite{ads} is avoided if $i_l \le N_f$ so that the theory does not have a runaway when $N_f \ge \frac{N_c}{k}$.  The development of a runaway coincides with when the dual gauge group($\tilde N_c)$ runs negative.

A self-dual point similar to SQCD exists for this theory.  Requiring that the dual has the same gauge group gives $N_f = \frac{2 N_c}{k}$.  The mesons are removed by the superpotential $W = X^{k+1} + \sum_{i+j=k-1} Q X^i \tilde Q Q X^j \tilde Q$.  As before, all of the terms in the superpotential are exactly marginal.  Unfortunately, this self-duality happens to be difficult to incorporate into a duality cascade.

A new self-dual point is realized for
\begin{center}
\be
\begin{tabular}{c|c|c}
&$SU(N)$&$SU(N-\delta)$\\
\hline
&&\\[-8pt]
$Q$ & $\fund$ & $\fund$ \\
$\tilde Q$ & $\overline \fund$ &  $\overline \fund$ \\
$X$ & $\adj$ &  
\end{tabular}
\ee
\end{center}
\bea
\nonumber
W = \lambda_1 \Tr X^{4} + \lambda_2 \Tr Q X^2 \tilde Q + \lambda_3 \Tr (Q \tilde Q)^2
\eea
Applying the above duality, we arrive at the theory
\begin{center}
\be
\begin{tabular}{c|c|c}
&$SU(2N-3\delta)$&$SU(N-\delta)$\\
\hline
&&\\[-8pt]
$q$ & $\fund$ & $\overline \fund$ \\
$\tilde q$ & $\overline \fund$ &  $\fund$ \\
$x$ & $\adj$ & \\
$M_0$ & & $1 + \adj$  \\
$M_1$ & & $1 + \adj$  \\
$M_2$ & & $1 + \adj$
\end{tabular}
\ee
\end{center}
\bea
\nonumber
W = - \lambda_1 \Tr x^{4} + \lambda_2 \Tr M_2 + \lambda_3 \Tr M_0^2 +  4 \frac{\lambda_1}{\mu^2} ( M_{2} q \tilde q + M_{1} q x \tilde q + M_{0} q x^2 \tilde q)
\eea
The F term for $M_2$ results in a vev $\langle q \tilde q \rangle = \mathds{1}_{N-\delta \times N-\delta}$.  The vev Higgses the gauge group while the remaining terms mass up various field.  After integrating out matter, the dual theory is
\begin{center}
\be
\begin{tabular}{c|c|c}
&$SU(N-2 \delta)$&$SU(N-\delta)$\\
\hline
&&\\[-8pt]
$q$ & $\fund$ & $\overline \fund$ \\
$\tilde q$ & $\overline \fund$ &  $\fund$ \\
$x$ & $\adj$ &
\end{tabular}
\ee
\end{center}
\bea
\nonumber
W = -\lambda_1 \Tr x^{4} - 4 \lambda_1 \Tr q x^2 \tilde q  - (2 \lambda_1 + \frac{\lambda_2^2}{4 \lambda_3})  \Tr (q \tilde q)^2
\eea
There is a self-dual point for $\delta = 0$ where the R charge of all fields is 1/2.  While the couplings do have a $1/\lambda_3$ component, it is not as simple as a coupling becoming its inverse.

Another self-dual fixed point is
\begin{center}
\be
\begin{tabular}{c|c|c}
&$SU(N)$&$SU(2N-\delta)$\\
\hline
&&\\[-8pt]
$Q$ & $\fund$ & $\fund$ \\
$\tilde Q$ & $\overline \fund$ &  $\overline \fund$ \\
$X$ & $\adj$ &  
\end{tabular}
\ee
\end{center}
\bea
\nonumber
W = \lambda_1 \Tr X^{3} + \lambda_2 \Tr Q X \tilde Q
\eea
Exactly at the self-dual point, the theory is connected to a free field theory.  Deforming epsilonically away from the free field theory, the theory flows to a Banks-Zaks fixed point.  The superpotential terms are relevant and the theory becomes strongly coupled.  Applying the above duality and integrating out matter, this theory is dual to
\begin{center}
\be
\begin{tabular}{c|c|c}
&$SU(N-\delta)$&$SU(2N-\delta)$\\
\hline
&&\\[-8pt]
$q$ & $\fund$ & $\overline \fund$ \\
$\tilde q$ & $\overline \fund$ &  $\fund$ \\
$x$ & $\adj$ &
\end{tabular}
\ee
\end{center}
\bea
\nonumber
W = - \lambda_1 \Tr x^{3} - 3 \lambda_1 \Tr q x \tilde q
\eea
This theory is IR free and after redefining the phase of $x$, there is a self-dual point for $\delta = 0$ and $\lambda_2 = 3 \lambda_1$.

\subsection{A chiral self-duality}
\label{Sec: SU chiral}

In this subsection, we present a chiral self-duality.  A duality for SU gauge groups with an antisymmetric tensor and a symmetric tensor was found in Ref.~\cite{Intriligator:1995ax}.  The electric theory is
\begin{center}
\be
\begin{tabular}{c|c|cc}
&$SU(N_c)$&$SU(N_f+8)$&$SU(N_f)$\\
\hline
&&\\[-8pt]
$Q$ & $\fund$ & $\fund$ & \\
$\tilde Q$ & $\overline \fund$ &  & $\fund$ \\
$A$ & $\asymm$ &  & \\
$\tilde S$ & $\overline \symm$ &  & 
\end{tabular}
\ee
\end{center}
\bea
\nonumber
W = \Tr (A \tilde S )^{2(k+1)}
\eea
Anomaly cancelation requires the additional 8 extra flavors of fundamental quarks.  This theory was demonstrated to be dual to 
\begin{center}
\be
\begin{tabular}{c|c|cc}
&$SU(\tilde N_c)$&$SU(N_f + 8)$&$SU(N_f)$\\
\hline
&&\\[-8pt]
$q$ & $\fund$ & $\overline \fund$ & \\
$\tilde q$ & $\overline \fund$ &  & $\overline \fund$ \\
$a$ & $\asymm$ &  & \\
$\tilde s$ & $\overline \symm$ &  & \\
$M_j = Q (\tilde S A)^j \tilde Q$ &  &  $\fund$ &  $\fund$ \\
$P_{r (\text{even)}} = Q (\tilde A S)^r \tilde S Q$ & & $\symm$ & \\
$P_{r (\text{odd)}} = Q (\tilde A S)^r \tilde S Q$ & & $\asymm$ & \\
$\tilde P_{r (\text{even)}} = \tilde Q A (\tilde S A)^r \tilde Q$ &  &  & $\asymm$ \\
$\tilde P_{r (\text{odd)}} = \tilde Q A (\tilde S A)^r \tilde Q$ &  &  & $\symm$
\end{tabular}
\ee
\end{center}
\bea
\nonumber
W = \Tr (a \tilde s )^{2(k+1)} + \sum_{j=0}^{2k+1} M_{2k+1-j} q (\tilde s a)^j \tilde q + \sum_{r=0}^{2k} P_{2k-r} q (\tilde s a)^r \tilde s q + \tilde P_{2k-r} \tilde q a (\tilde s a)^r \tilde q 
\eea
where the indices j runs from $0$ to $2k+1$ and r runs from $0$ to $2 k$ and $\tilde N_c = (4k+3)(N_f+4) -N_c$.  $P_r$($\tilde P_r$) are symmetric tensors when r is even (odd) and are antisymmetric tensors when r is odd (even).

There are a couple self-dualities that can be obtained using the above duality.  The one most relevant for constructing cascades is 
\begin{center}
\be
\label{Eq: SelfDual}
\begin{tabular}{c|c|cc}
&$SU(N)$&$SO(N-6-\delta)$&$SO(8)$\\
\hline
&&\\[-8pt]
$Q$ & $\fund$ & $\fund$ & \\
$Q'$ & $\fund$ & & $\fund$ \\
$\tilde Q$ & $\overline \fund$ & $\fund$ &\\
$A$ & $\asymm$ &  & \\
$\tilde S$ & $\overline \symm$ &  & 
\end{tabular}
\ee
\end{center}
\bea
\nonumber
W_1 = \Tr (A \tilde S )^2 + (Q \tilde Q)^2  + (Q' \tilde Q)^2  + (Q \tilde S A \tilde Q) + (Q' \tilde S Q)^2 + (Q' \tilde S Q')^2
\eea
which is dual to itself with gauge group $SU(N-2\delta)$.

This self-duality is different from the vector-like self-dualities of Sec.~\ref{Sec: SU adj} and App.~\ref{App: SOSP}.  At the self-dual point, the R charges of the fields are not 1/2.  Instead, a-maximization can be used to calculate their R charges\cite{Intriligator:2003jj}.  By tracking the flow as various couplings become relevant and part of the CFT\cite{Barnes:2004jj}, or by applying a-maximization subject to inequalities\cite{Hook:2012fp}, one arrives at the following R charges for the theory when $\delta=0$
\bea
\label{Eq: Rcharges}
R_Q = R_Q' = \frac{6 N +1}{18 N} \qquad R_{\tilde Q} = \frac{12N-1}{18N} \qquad R_A = \frac{6 N +1}{9 N} \qquad R_S = \frac{3 N - 1}{9 N}
\eea
Notice that these R charges are not 1/2 and do not go to it as N goes to infinity.


At infinite N, a new fixed point emerges.  For infinite N, the maximum is when all of the R charges are $1/2$.  From the point of view of a-maximization, in R charge space, as N increases a new maximum approaches the surface defined by the superpotential and anomaly free constraints.  Only at infinite N does this new maximum satisfy the constraints.  The previous maximum ceases to be a maximum when this new maximum appears.


\section{A $SU(N) \times SU(N)$ cascade}
\label{Sec: adjoint cascade}

Using these new self-dualities, we can build up several cascades involving two index tensors.  Many of the cascades are similar and only a few cascades will be presented each highlighting different features.  The first cascade we consider is a cascade involving adjoints.  We take the UV gauge theory
\begin{center}
\be
\begin{tabular}{c|cc|}
&$SU(N)$&$SU(N-\delta)$\\
\hline
&&\\[-8pt]
$Q$ & $\fund$ & $\fund$ \\
$\tilde Q$ & $\overline \fund$ &  $\overline \fund$ \\
$X_1$ & $\adj$ & \\
$X_2$ & & $\adj$ 
\end{tabular}
\ee
\end{center}
\bea
\label{Eq: super}
W = X_1^{4} + X_2^{4} + Q X_1^2 \tilde Q + Q X_2^2 \tilde Q + (Q \tilde Q)^2
\eea
There are many different UV fixed points and the flows that follow depend on the relative sizes of the different coefficients.  We describe one which behaves like the typical duality cascade.  For the starting UV fixed point, we take the self-dual $\delta=0$ fixed point.  We have the beta functions
\bea
\beta_{SU(N)} \sim \delta \qquad \beta_{SU(N-\delta)} \sim -\delta
\eea
where we have used that at the self-dual point the R charges are all 1/2.
Since the beta functions have opposite signs, we see that $SU(N-\delta)$ becomes more weakly coupled while $SU(N)$ becomes more strongly coupled.  Thus we arrive at the fixed point associated with the the $SU(N)$ gauge group.  Using a-maximization, we find that the most relevant quartic operator is the $X_1^4$ quartic.  Thus the theory flows near the fixed point presented in Sec.~\ref{Sec: SU adj}.

At the $SU(N)$ fixed point, the $SU(N-\delta)$ gauge coupling is irrelevant while the quartic $Q X_1^2 \tilde Q$ is relevant.  When this quartic coupling becomes large, the duality of Sec.~\ref{Sec: SU adj} shows us that this theory undergoes Higgsing effects.  Using the duality of Sec.~\ref{Sec: SU adj}, we see that this theory becomes a $SU(N-2\delta) \times SU(N - \delta)$ gauge theory.  Now the $SU(N-2\delta)$ gauge coupling is irrelevant while the $SU(N-\delta)$ gauge coupling is relevant leading to a cascade.  We are at the starting point with $N' = N - \delta$.  This cascade is much like the original duality cascade of KS except that the bottom of the cascade has new interesting physics possibilities.

This cascade highlights additional constraints that appear when building duality cascades.  Two self-dual points are not enough.  Dualizing one gauge group effects the superpotential terms that are used to build the other self-duality.  For example, the self-duality of Sec.~\ref{Sec: SU adj} has an effect on all of the terms in the superpotential shown in Eq.~\ref{Eq: super}.  The reason is that after duality, there exists an additional term $X_2^2 M_0$.  After $M_0$ gets integrated out, the superpotential terms $X_2^4$, $q X_2^2 \tilde q$ and $(q \tilde q)^2$ are all generated.  If these terms were not part of the self-duality of the $SU(N-\delta)$ then the duality cascade would not proceed.  

\subsection{IR dynamics}

We now consider the bottom of the cascade.  Take $N = m \delta + \epsilon$ with $m \in \mathbb{Z}$.  There are many cases and for simplicity they will be summarized here.  If $\epsilon = 0$, we have a goldstone mode and a free field in the IR.  If $0 < \epsilon < \frac{\delta}{2}$ the theory has a runaway that is stabilized by the superpotential and the theory obtains a mass gap.
If $\frac{\delta}{2} < \epsilon \le \frac{2 \delta}{3}$, there appears to be a long lived metastable SUSY breaking vacuum, though there is a singlet whose mass is sensitive to unknown Kahler potential corrections\footnote{This incalculability of the SUSY breaking vacuum is not present in the chiral cascade presented in Sec.~\ref{Sec: chiral cascade} or a vector-like cascade presented in App.~\ref{App: vector}.} and thus cannot be unambiguously shown to be non-tachyonic.  For $\frac{\delta}{2} < \epsilon < \delta$, a gauge symmetry breaking supersymmetric vacuum is found far away in field space from the origin.  

\subsubsection{$\epsilon = 0$: Free fields in the IR}
If $\epsilon=0$, the cascade ends with the gauge theory $SU(\delta) \times SU(2\delta)$.  
\begin{center}
\be
\begin{tabular}{c|cc|}
&$SU(\delta)$&$SU(2\delta)$\\
\hline
&&\\[-8pt]
$Q$ & $\fund$ & $\fund$ \\
$\tilde Q$ & $\overline \fund$ &  $\overline \fund$ \\
$X_1$ & $\adj$ & \\
$X_2$ & & $\adj$ 
\end{tabular}
\ee
\end{center}
\bea
\nonumber
W = X_1^{4} + X_2^{4} + Q X_1^2 \tilde Q + Q X_2^2 \tilde Q + (Q \tilde Q)^2
\eea
Dualizing the $SU(2\delta)$ gauge theory yields a $SU(\delta) \times SU(\delta)$ gauge theory.  The dual quark vevs now break $U(1)_B$ yielding a goldstone mode.  Additionally the singlet meson $M_1$ remains massless.  The remaining modes mass up and the remaining $SU(\delta)$ gauge group is left with just an adjoint which confines.  Thus the IR theory is simply the goldstone mode and the singlet meson $M_1$.

\subsubsection{$0 < \epsilon < \frac{\delta}{2}$: Confinement}
For $0 < \epsilon < \frac{\delta}{2}$, the cascade ends in the theory shown below
\begin{center}
\be
\label{Eq: laststep}
\begin{tabular}{c|cc|}
&$SU(\epsilon)$&$SU(\epsilon + \delta)$\\
\hline
&&\\[-8pt]
$Q$ & $\fund$ & $\fund$ \\
$\tilde Q$ & $\overline \fund$ &  $\overline \fund$ \\
$X_1$ & $\adj$ & \\
$X_2$ & & $\adj$ 
\end{tabular}
\ee
\end{center}
\bea
\nonumber
W = X_1^{4} + X_2^{4} + Q X_1^2 \tilde Q + Q X_2^2 \tilde Q + (Q \tilde Q)^2
\eea
Given these choice of parameters, the $SU(\epsilon + \delta)$ theory has a runaway as described in Sec.~\ref{Sec: SU adj}.  We give an explicit demonstration of this effect when $\epsilon=\frac{\delta-2}{2}$.

  To find an explicit form of the runaway, we start with the situation where $SU(\epsilon+\delta)$ confines with a superpotential and integrate out a flavor.  The confining case was worked out in Ref.~\cite{Csaki:1998fm} and the confining superpotential is shown in Eq.~\ref{Eq: confine}.  Integrating out a flavor, one arrives at the superpotential
\bea
W_{\text{dyn}} &\sim& \frac{1}{(\text{det} M_2)^{\frac{9}{2}}} ( (\text{det} M_2)^2 (M_0 \text{cof} M_2) ) +  (\text{det} M_2) (M_1 \text{cof} M_2)^2 ) ) \\
\text{cof} M &=& \frac{\partial \text{det} M}{\partial M_{ij}}
\eea
The runaway is stabilized by the two superpotential terms $M_2$ and $M_0^2$ and the theory obtains a mass gap.  If we have $\epsilon=\frac{\delta}{2}$, presumably the $SU(\epsilon + \delta)$ confines with a quantum moduli space, though the details have not yet been worked out.

\subsubsection{$\frac{\delta+1}{2} \le \epsilon < \delta$: SUSY breaking and restoration}

Next we consider the case when $\frac{\delta+1}{2} \le \epsilon < \delta$.  As before, the last step of the cascade is shown in Eq.~\ref{Eq: laststep}.  
The $SU(\epsilon)$ gauge coupling is irrelevant while the $SU(\epsilon + \delta)$ gauge coupling is relevant.  We will show that a long lived meta-stable vacuum potentially\footnote{The mass of one of the singlets is incalculable.  This problem is not present in cascades presented in Sec.~\ref{Sec: chiral cascade} or App.~\ref{App: vector}.} exists when $\Lambda_{SU(\epsilon)} \gg \Lambda_{SU(\epsilon + \delta)}$.
Before the  $SU(\epsilon + \delta)$'s gauge coupling becomes strong, the quartic interactions are all irrelevant.  After $SU(\epsilon + \delta)$ becomes strongly coupled,  a-maximization can be used to show that the quartic operators are relevant with $X_2^4$ being the most relevant operator.
We dualize the $SU(\epsilon+\delta)$ gauge group and obtain the IR theory
\begin{center}
\be
\begin{tabular}{c|cc|}
&$SU(\epsilon)$&$SU(2\epsilon-\delta)$\\
\hline
&&\\[-8pt]
$q$ & $\overline \fund$ & $\fund$ \\
$\tilde q$ & $\fund$ &  $\overline \fund$ \\
$X_1$ & $\adj$ & \\
$x_2$ & & $\adj$ \\
$M_0$ & $1+\adj$ & \\
$M_1$ & $1+\adj$ & \\
$M_2$ & $1+\adj$ & 
\end{tabular}
\ee
\end{center}
\bea
\nonumber
W = X_1^{4} + x_2^{4} + X_1^2 M_0 + M_2 + M_0^2 + M_2 q \tilde q + M_1 q x_2 \tilde q + M_0 q x_2^2 \tilde q
\eea
The F term for $M_2$ requires that $q \tilde q \sim \mathds{1}_{\epsilon \times \epsilon}$, however the rank of $\text{Rank}(q \tilde q) = \text{min}(\epsilon,2\epsilon-\delta) = 2\epsilon-\delta < \epsilon$.   So it is impossible to satisfy the F term for $M_2$.  Thus we see that we have a candidate SUSY breaking vacuum. 

We first study the candidate SUSY breaking vacuum.  In the UV, all of the perturbations are irrelevant so the deformations are all small in value and that perturbation theory is reliable.  We decompose our theory around the vacuum
\bea
q  = \begin{pmatrix}
  \mathds{1} + \chi_+ + \chi_-  \\
  \rho_+ + \rho_-
 \end{pmatrix} \qquad \tilde q  = \begin{pmatrix}
  \mathds{1} + \chi_+ - \chi_-  \\
  \rho_+ - \rho_-
 \end{pmatrix} \\
 X_1 = \begin{pmatrix}
  \alpha & \beta  \\
  \tilde \beta & \gamma
 \end{pmatrix} \qquad M_{0,1,2} = \begin{pmatrix}
  Y_{0,1,2} & Z_{0,1,2}  \\
  \tilde Z_{0,1,2} & \Phi_{0,1,2}
 \end{pmatrix} 
\eea
As in the ISS scenario, the fields inside of the $q$,$\tilde q$, and $M_2$ fields obtain either a tree level mass, are eaten by gauge bosons or obtain a positive 1-loop mass.  $M_0$ , $Y_1$ and $x_2$ have a supersymmetric tree level mass.  All fields in $M_1$, $M_2$, $X_1$, $x_2$ charged under either gauge group receive a positive 2-loop mass from gauge mediation.  The only matter field which is not stabilized at this point is the singlet part of $M_1$.  Being coupled via an irrelevant operator, its mass cannot reliably be determined\cite{Intriligator:2008fe}.  Incalculable Kahler potential corrections to the mass of $M_1$ are of the form
\bea
V \sim \frac{q q^\dagger M_1^2}{\Lambda^2}
\eea
The loop level corrections to its mass from the superpotential are
\bea
m_\text{superpotential} \sim \frac{|\langle q \rangle|^2}{ (16 \pi^2)^L \Lambda^2}
\eea
where $L$ is the loop level.  The incalculable Kahler potential contributions to the mass of $M_1$ are larger than the calculable loop level corrections.  
Thus while it appears that there is likely a SUSY breaking vacuum, its stability cannot be reliably determined.

In Ref.~\cite{Intriligator:2006dd}, a supersymmetric vacuum were found at large field values.  For $\epsilon=\frac{\delta+1}{2}$, $SU(\epsilon + \delta)$ confines with the superpotential\cite{Csaki:1998fm} 
\bea
\label{Eq: confine}
W_{\text{dyn}} &\sim& \tilde q M_2 q + (\text{det} M_2)^2 (M_0 \text{cof} M_2) + (\text{det} M_2) (M_1 \text{cof} M_2)^2 
\eea
One can explicitly show that this dynamical superpotential allows for the existence of a supersymmetric vacuum at large field values. If $\epsilon>\frac{\delta+1}{2}$ gaugino condensation at large field values allows for the existence of a supersymmetric vacuum.  Consider large field values of the singlet in $M_2$ and giving the adjoint the vev\footnote{App.~\ref{App: example} shows how taking symmetry breaking vevs can be used to rederive the well known ADS superpotential.}
\bea
x_2 = x \left( \begin{array}{ccc}
\mathds{1}_{2 \epsilon - \delta - 1 \times 2 \epsilon - \delta - 1} & 0 \\
0 & -2 \epsilon + \delta + 1 \end{array} \right)
\eea
The vev in the adjoint breaks the gauge group down to $SU(2 \epsilon - \delta - 1) \times U(1)$ while $M_2$ lifts all of the matter for this unbroken gauge group.  The IR matter content of  $SU(2 \epsilon - \delta - 1)$ consists of a single adjoint.  This adjoint obtains a mass from the quartic so that the final dynamical scale corresponding to gaugino condensation is
\bea
\Lambda_{SU(2 \epsilon - \delta - 1)}^{ 6 \epsilon - 3 \delta - 3} \sim M_2^\epsilon \Lambda_{SU(2 \epsilon - \delta)}^{\epsilon - \delta + 1} x^{4 \epsilon - 2 \delta - 4}
\eea
Finally, the linear and quartic term in conjunction with gaugino condensation has a superpotential
\bea
W &\supset& (M_2^\epsilon \Lambda_{SU(2 \epsilon - \delta)}^{\epsilon - \delta + 1} x^{4 \epsilon - 2 \delta - 4})^{1/(2 \epsilon - \delta - 1)} + \mu^2 M_2 + \frac{x^4}{\Lambda_{SU(2 \epsilon - \delta)}}
\eea
The linear term was irrelevant in the UV so that we have $\mu \sim \frac{\Lambda_{SU(\epsilon+\delta)}^2}{\Lambda_{SU(\epsilon)}} \ll \Lambda_{SU(2 \epsilon - \delta)} \sim \Lambda_{SU(\epsilon+\delta)}$, given our initial assumption that the original dynamical scales were far apart.
We see that we have a supersymmetric vacuum at
\bea
M_2 &\sim& \Lambda_{SU(2 \epsilon - \delta)} (\frac{\mu}{\Lambda_{SU(2 \epsilon - \delta)}})^\frac{4 \epsilon - 2\delta}{\delta} \sim \mu (\frac{\mu}{\Lambda_{SU(2 \epsilon - \delta)}})^\frac{4\epsilon-3\delta}{\delta} \\
x &\sim& \Lambda_{SU(2 \epsilon - \delta)} (\frac{\mu}{\Lambda_{SU(2 \epsilon - \delta)}})^\frac{\epsilon}{\delta} \sim \mu (\frac{\mu}{\Lambda_{SU(2 \epsilon - \delta)}})^\frac{\epsilon-\delta}{\delta}
\eea
We had assumed that the vev of $x_2$ broke the gauge group down to $SU(2\epsilon - \delta -1)$, but the scaling of these solutions are independent of which gauge group we broke $SU(2\epsilon-\delta)$ down to.  We are studying the range $\frac{\delta+1}{2} \le \epsilon < \delta$.  In this range the supersymmetric vacuum always obeys $M_2, x \ll \Lambda_{SU(2 \epsilon - \delta)}$.  Thus our supersymmetric vacuum is under control.  For it to be long lived, we need $M_2$ or $x \gg \mu$.  We find that $x \gg \mu$ for all values in the range of interest while $M_2 \gg \mu$ if $\epsilon < \frac{3 \delta}{4}$.  As this vacuum goes to infinity as $\epsilon \rightarrow 0$, our meta-stable SUSY breaking vacuum, if it exists, is long lived.

If $\frac{\delta}{2} < \epsilon \le \frac{2 \delta}{3}$, $SU(2 \epsilon - \delta)$ is IR free while if $\frac{2 \delta}{3} < \epsilon < \delta$, $SU(2 \epsilon - \delta)$ is a CFT.  If the gauge group is a CFT, then the perturbative analysis conducted before around the SUSY breaking vacuum is not reliable and conformal perturbation theory would need to be applied.  The SUSY breaking vacuum is always far away in field space from the origin.

\section{A chiral cascade}
\label{Sec: chiral cascade}

One can build a cascade from the chiral self-dual theory presented in Sec.~\ref{Sec: SU chiral}.  Gauging the $SO(N-6)$ gauge group, we see that at infinite N, the theory is self-dual under both duality transformations of the $SU$ and $SO$ gauge groups.  Assume we start with the gauge group $SU(N) \times SO(N-6-\delta)$ with $N,\delta \gg 1$.  The gauge theory and superpotential of the starting theory is
\begin{center}
\be
\begin{tabular}{c|cc|c}
&$SU(N)$&$SO(N-6-\delta)$&$SO(8)$\\
\hline
&&\\[-8pt]
$Q$ & $\fund$ & $\fund$ & \\
$Q'$ & $\fund$ & & $\fund$ \\
$\tilde Q$ & $\overline \fund$ & $\fund$ &\\
$A$ & $\asymm$ &  & \\
$\tilde S$ & $\overline \symm$ &  & 
\end{tabular}
\ee
\end{center}
\bea
\nonumber
W = \Tr (A \tilde S )^2 + (Q \tilde Q)^2  + (Q' \tilde Q)^2  + (Q \tilde S A \tilde Q) + (Q' \tilde S Q)^2 + (Q' \tilde S Q')^2
\eea
Starting near the infinite N fixed point, finite N effects render the $SO$ gauge coupling irrelevant and the $SU$ gauge coupling relevant.  Thus we dualize the $SU$ gauge group to obtain the theory $SU(N-2\delta) \times SO(N-6-\delta)$.  Now the $SO$ gauge group's gauge coupling becomes stronger while the $SU$ gauge group becomes weaker.  Applying Seiberg duality to the $SO$ gauge group\cite{Seiberg:1994pq,Intriligator:1995id} yields the final theory $SU(N-2\delta) \times SO(N-3\delta + 10)$.  Thus we see that we have a cascade which exhibits the effect that
\bea
N' = N-2 \delta \qquad \delta' = \delta - 16
\eea
Thus the cascade slows down as it approaches the IR!  If the cascade does not end before $\delta$ runs negative, then repeated applications of duality does not reduce the rank of the gauge groups.

After n steps of the cascade we see that 
\bea
N(n) = N - 2 n \delta + 16 n(n-1) \qquad \delta(n) = \delta - 16 n
\eea
The minimum occurs at 
\bea
n_c = \frac{\delta+8}{16} \qquad N_c = N - (\frac{\delta+8}{4})^2
\eea
If we want the cascade to end, it must do so before repeated dualities start increasing the gauge group.  Large $\delta$ ensures many applications of the duality operation.

The IR dynamics of this cascade are many and varied.  In what follows, only a few examples are taken to highlight the new possibilities.  The duality can end with the $SO$ gauge group.  As an example take the cascade ending with $N = 3 \delta - 10$ and $\delta > 10$.  The $SU$ gauge group dualizes to a $SU(\delta-10) \times SO(2\delta-16)$ gauge group.  Now we notice that the $SO$ gauge group confines.  There are two physically distinct branches, one with an ADS superpotential and one without.  We take the branch with no ADS superpotential.  The mesons have explicit mass terms and in the IR we are left with the theory
\begin{center}
\be
\begin{tabular}{c|c|c}
&$SU(\delta-10)$&$SO(8)$\\
\hline
&&\\[-8pt]
$Q'$ & $\fund$ & $\fund$ \\
$A$ & $\asymm$  & \\
$\tilde S$ & $\overline \symm$ &  
\end{tabular}
\ee
\end{center}
\bea
\nonumber
W = \Tr (A \tilde S )^2 + (Q' \tilde S Q')^2
\eea
So we have an interacting CFT in the IR.  

If instead the $SU$ gauge group ends the cascade, it does so with supersymmetry breaking.
\begin{center}
\be
\begin{tabular}{c|cc|c}
&$SU(2N-6-3\delta)$&$SO(N-6-\delta)$&$SO(8)$\\
\hline
&&\\[-8pt]
$q$ & $\fund$ & $\fund$ & \\
$q'$ & $\fund$ & & $\fund$ \\
$\tilde q$ & $\overline \fund$ & $\fund$ &\\
$a$ & $\asymm$ &  & \\
$\tilde s$ & $\overline \symm$ &  & \\
$M_0 = Q \tilde Q$ &  & $\fund^2$  & \\
$M'_0 = Q' \tilde Q$ &  & $\fund$  & $\fund$ \\
$M_1 = Q \tilde S A \tilde Q$ &  & $\fund^2$  & \\
$M'_1 = Q' \tilde S A \tilde Q$ &  & $\fund$  & $\fund$ \\
$P = Q \tilde S Q$ &  & $\symm$  & \\
$P' = Q' \tilde S Q$ &  & $\fund$  & $\fund$ \\
$P'' = Q' \tilde S Q'$ &  &  & $\symm$ \\
$\tilde P = Q \tilde S Q$ &  & $\asymm$  & 
\end{tabular}
\ee
\end{center}
\bea
\nonumber
W &=& \Tr (a \tilde s)^2 + M_0^2  + M'^2_0  + M_1 + P'^2 + P''^2 + q \tilde q M_1 + q' \tilde q M'_1 + q \tilde s a \tilde q  M_0 \\
\nonumber
&+& q' \tilde s a \tilde q M'_0 + \tilde q a \tilde q \tilde P + q \tilde s q P + q \tilde s q' P' + q' \tilde s q' P''
\eea
Taking $N < 2 \delta$, we find that supersymmetry is broken through the rank condition.  As before, the F term for $M_1$,$\langle q \tilde q \rangle = \mathds{1}$, cannot be satisfied as the rank of $\langle q \tilde q \rangle$ is not large enough.  As in the previous example, all of the bosons have positive 1-loop or 2-loop masses.  Two singlets $P_1$ and $P_2$ deserve special attention.  We decompose
\bea
P = \begin{pmatrix}
  P_1 \mathds{1}_{2N-6-3\delta \times 2N-6-3\delta} & 0  \\
  0 & P_2 \mathds{1}_{2\delta-N \times 2\delta-N}
 \end{pmatrix} 
\eea
The singlet which gave problems in the previous example is the sum $P_1 + P_2$.  The vevs of $q$ and $\tilde q$ result in a mass term between $P_1$ and $\tilde s$.  Thus the remaining singlet is $P_2$.  Notice that while $P_2$ is a singlet under the surviving $SO(2N-6-3\delta)$ symmetry, it is not a singlet under the original $SO(N-6-\delta)$ gauge symmetry.  Thus it receives a 2-loop positive mass from Higgsed gauge mediation\cite{Gorbatov:2008qa}.  Thus we find that all bosons are stabilized so that we have a bonafide SUSY breaking vacuum.

  To find supersymmetric vacua at large field values, we give large vevs to $a$ and $\tilde s$.  A D flat direction is 
\bea
\langle a \rangle =  \begin{pmatrix}
  a_1 \sigma_2 &  &  \\
   & \ddots  &  \\
   &  & a_n \sigma_2
 \end{pmatrix}  &\qquad&
\langle \tilde s \rangle = \begin{pmatrix}
  s_1 \mathds{1} &  &  \\
   & \ddots  &  \\
   &  & s_n \mathds{1}
 \end{pmatrix} \\
 |a_i|^2-|s_i|^2 &=& \text{constant}
\eea
For simplicity, we'll explore the direction parameterized by $a_i = s_i = x$.  This vev breaks the gauge group down to $U(\lfloor \frac{2N-6-3\delta}{2} \rfloor)$.  The quarks decompose into $2N-4-2\delta$ flavors.  These flavors can be given a mass by going out in field space for the mesons $M_1$ and $P''$.  Now the $U(\lfloor \frac{2N-6-3\delta}{2} \rfloor)$ has no matter and gaugino condensation occurs.  For simplicity, we assume $\delta$ is even so that the floor function can be ignored.
\bea
\Lambda_{SU(N-3-3\delta/2)}^{3N-9-9\delta/2} = M_1^{2N-12-2\delta} \Lambda^{5N-25-17\delta/2}_{SU(2N-6-3\delta)} x^{-4N+20+6\delta} P''^8
\eea
Thus the superpotential is 
\bea
W \supset (M_1^{2N-12-2\delta} \Lambda^{5N-25-17\delta/2}_{SU(2N-6-3\delta)} x^{-4N+20+6\delta})^{1/(N-3-3\delta/2)} + \frac{x^4}{\Lambda_{SU(2N-6-3\delta)}} + \mu_1^2 M_1 + \mu_2 P''^2
\eea
For simplicity, we will assume that $\mu_1 \sim \mu_2  = \mu$.  By adjusting the ratio of the UV dynamical scales we can live in the region $\mu \ll \Lambda_{SU(2N-6-3\delta)}$.  The solution for the SUSY preserving vacuum is then
\bea
P &\sim& \Lambda_{SU(2N-6-3\delta)} (\frac{\mu}{\Lambda_{SU(2N-6-3\delta)}})^\frac{2N-5\delta/2-10}{\delta} \sim \mu (\frac{\mu}{\Lambda_{SU(2N-6-3\delta)}})^\frac{2N-7\delta/2-10}{\delta} \\
M_1 &\sim& \Lambda_{SU(2N-6-3\delta)} (\frac{\mu}{\Lambda_{SU(2N-6-3\delta)}})^\frac{4N-6\delta-20}{\delta} \sim \mu (\frac{\mu}{\Lambda_{SU(2N-6-3\delta)}})^\frac{4N-7\delta-20}{\delta} \\
x &\sim& \Lambda_{SU(2N-6-3\delta)} (\frac{\mu}{\Lambda_{SU(2N-6-3\delta)}})^\frac{N-\delta-5}{\delta} \sim \mu (\frac{\mu}{\Lambda_{SU(2N-6-3\delta)}})^\frac{N-2\delta-5}{\delta} 
\eea
We had assumed that the gauge group was broken down to $U(\lfloor \frac{2N-6-3\delta}{2} \rfloor)$.
As before, if a different vev was chosen to give a different symmetry breaking pattern, the SUSY vacuum has the same scaling behavior.
We are working in the range $\frac{3\delta}{2}+3 < N < 2\delta$. 
 As long as $2N-6-3\delta > 4$, we have $P,M_1,x \ll \Lambda$ so that our vacuum is under control.  If $2N-6-3\delta \le 4$, the vev of $M_1$ goes beyond the UV cutoff.  
 We always have $x \gg \mu$ so that we have a long lived meta-stable SUSY breaking vacuum.

When the IR gauge group is $\le 4$, there are special cases.  These correspond to the cases mentioned before where the vev of $M_1$ goes beyond the UV cutoff.  It is expected that a supersymmetric vacuum exist for these cases, but using our present understanding of the chiral gauge theory, it cannot be shown to be the case.  For IR gauge group $SU(1)$, we have a confining theory where instantons presumably generate a dynamical superpotential which causes a supersymmetric vacuum to exist. For $SU(2)$ and $SU(3)$, giving a vev to the two index tensors does not break the gauge group down to a non-abelian gauge group and instead down to a $U(1)$ gauge theory.  Again, we would expect instanton effects to generate a dynamical superpotential that restores supersymmetry\footnote{This intuition is from App.~\ref{App: example} where instantons generate the ADS superpotential when the gauge group is broken to $U(1)$.}.  Finally for a gauge group $SU(4)$, we have the supersymmetric vacuum at $M_1 \sim \Lambda$ so that the vacuum is not under control.

\section{A free field theory based cascade}
\label{Sec: free cascade}

We construct a cascade based on a  self-dual free fixed point.  Rather than being deformed into a weakly coupled Banks-Zaks fixed point, it becomes strongly coupled due to relevant superpotential interactions.  The starting point of this next cascade is 
\begin{center}
\be
\label{Eq: free}
\begin{tabular}{c|cc|}
&$SU(2N-\delta)$&$SU(N)^4$\\
\hline
&&\\[-8pt]
$Q_i + \tilde Q_i$ & $\fund$ & $\fund$ \\
$X$ & $\adj$ & \\
$X_i$ & & $\adj$ 
\end{tabular}
\ee
\end{center}
\bea
\nonumber
W = X^{3} + X_i^{3} + Q_i X \tilde Q_i + Q_i X_i \tilde Q_i
\eea
where the notation $SU(N)^4$ indicates 4 different $SU(N)$ gauge groups all with bifundamentals connecting them to the $SU(2N)$ gauge theory.  For simplicity, consider the scenario where all 4 $SU(N)$ gauge groups are related by a $\mathbb{Z}_4$ symmetry.   Applying the cubic duality of Sec.~\ref{Sec: SU adj}, we see that we have the cascade, $SU(2N-\delta) \times SU(N)^4 \rightarrow SU(2N-\delta) \times SU(N-\delta)^4  \rightarrow SU(2N-3\delta) \times SU(N-\delta)^4$ so that after two applications of the duality, we arrive back at the original situation with $N' = N - \delta$.

As before, up to issues with an incalculable singlet, the theory can have meta-stable SUSY breaking in the IR.  If $N = m \delta + \epsilon$ with $m \in \mathbb{Z}$, for $0 \lesssim \epsilon \lesssim \frac{\delta}{6}$ ($\frac{\delta}{2} \lesssim \epsilon \lesssim \frac{2 \delta}{3}$) the $SU(2N-\delta)$ $( SU(N)^4 )$ gauge group develops a runaway that is stabilized by the superpotential and the theory develops a mass gap.  For $\frac{\delta}{6} \lesssim \epsilon \lesssim \frac{\delta}{2}$ ($\frac{2 \delta}{3} \lesssim \epsilon \lesssim \delta$), the $SU(2N-\delta)$ ($SU(N)^4$) gauge group potentially has metastable SUSY breaking.  For simplicity, we have ignored edge cases. 

The runaway here is stabilized in a different manner than is typically seen.  Assume that we're at the bottom of the cascade and the $SU(N)^4$ gauge group has a run away\cite{Csaki:1998fm}.  The starting theory is shown in Eq.~\ref{Eq: free}  For simplicity, take $N = \frac{2 \delta -1}{3}$ so that the low energy theory is
\begin{center}
\be
\begin{tabular}{c|c|}
&$SU(2N-\delta)$\\
\hline
&\\[-8pt]
$X$ & $\adj$ \\
$M_{0,i}=Q_i \tilde Q_i$ & $1+\adj$ \\
$M_{1,i}=Q_i X_i \tilde Q_i$ & $1+\adj$ 
\end{tabular}
\ee
\end{center}
\bea
\nonumber
W = X^{3} + M_{1,i} + M_{0,i} X + \frac{M_{0,i} \text{cof} M_{1,i}}{(\text{det} M_{1,i})^3}
\eea 
The F term for the singlet part of $M_{0,i}$ still sets $M_{1,i} \rightarrow \infty$.  So one might expect that there is still a runaway.  However, from the original electric theory, we can show that there exists another branch where SUSY vacua exists.  

Recall the original argument for the runaway in the electric theory.  In Sec.~\ref{Sec: SU adj}, a vev was given to $X_i$ and it was observed that obtaining a standard ADS superpotential was unavoidable.  However, the new superpotential has the coupling $Q_i X_i \tilde Q_i$ so that giving $X_i$ a vev also masses up all of the flavors and gaugino condensation occurs instead of an ADS runaway.  As before, give $X_i$ a vev
\bea
X_i = x_i \left( \begin{array}{ccc}
\mathds{1}_{N - 1 \times N - 1} & 0 \\
0 & -N+1 \end{array} \right)
\eea
This vev breaks the gauge group down to $SU(N-1)$ and masses up both the adjoint and Q.  Thus we have a new dynamical scale which is
\bea
 \Lambda_{N-1}^{3N-3} = \Lambda_{N}^{\delta} x^{3N-\delta-3}
\eea
Thus after gaugino condensation, the superpotential is 
\bea
W \supset ( \Lambda_{N}^{\delta} x_i^{3N-\delta-3})^{1/(N-1)} + \lambda x_i^3
\eea
Solving for the vev $x_i$, we find that there is the solution 
\bea
x_i \sim \lambda^\frac{1-N}{\delta} \Lambda = \lambda^\frac{2(2-\delta)}{3 \delta} \Lambda
\eea
So we see that we can trust our vacuum as long as $\lambda \ll 1$ so that $x_i \gg \Lambda$.  As SUSY theories do not have phases transitions when varying parameters, we expect that there exists a SUSY vacuum even when $\lambda \gtrsim 1$.  Our runaway is cured by a gauge symmetry breaking SUSY vacuum!  This vacuum is seen in the electric theory rather than the magnetic theory as the vevs are large.

\section{Conclusion}
\label{Sec: conclusion}

In this article, we presented new self-dualities and  several different cascades with interesting physics in the IR.  These cascades all involved two index tensors and their IR physics included confinement, CFTs, and meta-stable supersymmetry breaking.
The self-dualities used exhibited Higgsing effects and utilized cubic, quartic and sextic operators.
A chiral cascade was constructed which slowed down in the IR and had its self-dual point at infinite N.

The gravity duals of these cascades would be very interesting.  As the field theory is under control, it would be interesting to check what do the metastable vacuum correspond to in the gravity dual.  As the gravity dual involves a large N limit where the dynamical scales and quartic couplings are all roughly equal, the meta-stable SUSY breaking vacuum and the SUSY vacuum would be exponentially close and the meta-stable vacuum would likely cease to be stable.

The cascades presented in this paper do not have moduli spaces where $\mathcal{N}=4$ $SU(N)$ gauge groups appear so it is unlikely that these cascades appear as D3 branes at a singularity.  Finding different cascades which admit a brane realization as D3 branes at a conifold would prove very enlightening.

While cascades other than the original KS construction have been proposed, they have mainly been orientifolds of the original picture.  In the confining region, the gravity duals of these new cascades would geometerize confinement in the presence of two-index tensors.  It would be interesting to compare with the original solution for any similarities.

%
%
%
%

\section*{Acknowledgements}
 AH would like to thank G. Torroba for early collaboration in the project and N. Seiberg for helpful discussions. AH would like to thank G. Torroba for helpful comments on the draft.  AH is supported by the Department of Energy under contract DE-SC0009988.

\appendix

\section{More self-dualities}
\label{App: SOSP}

Many new self-dualities can be rederived using Higgsing effect.  In this appendix, we present additional self-dualities which are useful in building duality cascades.

\subsection{$SU(N)$ with a flavor of symmetrics}
\label{Sec: SU sym}

A duality for SU gauge groups with a flavor of symmetric tensors and $N_f$ flavors was found in Ref.~\cite{Intriligator:1995ax}.  The electric theory is
\begin{center}
\be
\begin{tabular}{c|c|ccc}
&$SU(N_c)$&$SU(N_f)_L$&$SU(N_f)_R$&$U(1)_R$\\
\hline
&&\\[-8pt]
$Q$ & $\fund$ & $\fund$ & & $1-\frac{N_c-2k}{(k+1) N_f}$\\
$\tilde Q$ & $\overline \fund$ &  & $\fund$ & $1-\frac{N_c-2k}{(k+1) N_f}$\\
$S$ & $\symm$ &  & & $\frac{1}{k+1}$ \\
$\tilde S$ & $\overline \symm$ &  & & $\frac{1}{k+1}$ 
\end{tabular}
\ee
\end{center}
\bea
\nonumber
W = \Tr (S \tilde S )^{k+1}
\eea
This theory was demonstrated to be dual to 
\begin{center}
\be
\begin{tabular}{c|c|ccc}
&$SU(\tilde N_c)$&$SU(N_f)_L$&$SU(N_f)_R$&$U(1)_R$\\
\hline
&&\\[-8pt]
$q$ & $\fund$ & $\overline \fund$ & & $1-\frac{\tilde N_c-2k}{(k+1) N_f}$ \\
$\tilde q$ & $\overline \fund$ &  & $\overline \fund$ & $1-\frac{\tilde N_c-2k}{(k+1) N_f}$  \\
$s$ & $\symm$ &  & & $\frac{1}{k+1}$ \\
$\tilde s$ & $\overline \symm$ &  & & $\frac{1}{k+1}$ \\
$M_j = Q (\tilde S S)^j \tilde Q$ &  &  $\fund$ &  $\fund$ & $2-\frac{2(N_c-2k)}{(k+1) N_f} + \frac{2j}{k+1}$ \\
$P_r = Q (\tilde S S)^r \tilde S Q$ & & $\symm$ &  $\fund$ & $2-\frac{2(N_c-2k)}{(k+1) N_f} + \frac{2r+1}{k+1}$ \\
$\tilde P_r = \tilde Q S (\tilde S S)^r \tilde Q$ &  & &  $\symm$& $2-\frac{2(N_c-2k)}{(k+1) N_f} + \frac{2r+1}{k+1}$ 
\end{tabular}
\ee
\end{center}
\bea
\nonumber
W = \Tr (s \tilde s )^{k+1} + \sum_{j=0}^k M_{k-j} q (\tilde s s)^j \tilde q + \sum_{r=0}^{k-1} P_{k-r-q} q (\tilde s s)^r \tilde s q + \tilde P_{k-r-q} \tilde q s (\tilde s s)^r \tilde q 
\eea
where the indices j runs from $0$ to $k$ and r runs from $0$ to $k-1$ and $\tilde N_c = (2k+1)N_f + 4k -N_c$.   Applying the same logic as in Sec.~\ref{Sec: SU adj}, we find that a runaway develops if $N_f < \frac{N_c-4k}{2k+1}$.   As before, this runaway occurs when the dual gauge group($\tilde N_c$) runs negative.

Using this duality, one can show that the theory
\begin{center}
\be
\begin{tabular}{c|c|c}
&$SU(N)$&$SU(N-2-\delta)$\\
\hline
&&\\[-8pt]
$Q$ & $\fund$ & $\fund$  \\
$\tilde Q$ & $\overline \fund$ &  $\fund$ \\
$S$ & $\symm$ &  \\
$\tilde S$ & $\overline \symm$  & 
\end{tabular}
\ee
\end{center}
\bea
\nonumber
W = \Tr (S \tilde S )^2 + (Q \tilde Q )^2 + \Tr (Q \tilde S S \tilde Q)
\eea
is dual to itself with new gauge group $SU(N-2\delta)$.  Thus there is a self-duality for $\delta=0$.  At the self-dual point, the R charge of all fields is 1/2.

\subsection{$SU(N)$ with a flavor of antisymmetric tensors}
\label{Sec: SU asym}

A duality for SU gauge groups with a flavor of antisymmetric tensors and $N_f$ flavors was found in Ref.~\cite{Intriligator:1995ax}.  The electric theory is 
\begin{center}
\be
\begin{tabular}{c|c|ccc}
&$SU(N_c)$&$SU(N_f)_L$&$SU(N_f)_R$&$U(1)_R$\\
\hline
&&\\[-8pt]
$Q$ & $\fund$ & $\fund$ & & $1-\frac{N_c+2k}{(k+1)N_f} $\\
$\tilde Q$ & $\overline \fund$ &  & $\fund$ & $1-\frac{N_c+2k}{(k+1)N_f}$\\
$A$ & $\asymm$ &  & & $\frac{1}{k+1}$\\
$\tilde A$ & $\overline \asymm$ &  & & $\frac{1}{k+1}$
\end{tabular}
\ee
\end{center}
\bea
\nonumber
W = \Tr (A \tilde A )^{k+1}
\eea
There are also two additional $U(1)$ symmetries.  This theory was demonstrated to be dual to 
\begin{center}
\be
\begin{tabular}{c|c|ccc}
&$SU(\tilde N_c)$&$SU(N_f)_L$&$SU(N_f)_R$&$U(1)_R$\\
\hline
&&\\[-8pt]
$q$ & $\fund$ & $\overline \fund$ & & $1-\frac{\tilde N_c+2k}{(k+1)N_f} $\\
$\tilde q$ & $\overline \fund$ &  & $\overline \fund$ & $1-\frac{\tilde N_c+2k}{(k+1)N_f} $\\
$a$ & $\asymm$ &  & & $\frac{1}{k+1}$ \\
$\tilde a$ & $\overline \asymm$ &  & & $\frac{1}{k+1}$\\
$M_j = Q (\tilde A A)^j \tilde Q$ &  &  $\fund$ &  $\fund$ & $2-\frac{2(N_c+2k)}{(k+1)N_f} + \frac{2j}{k+1}$ \\
$P_r = Q (\tilde A A)^r \tilde A Q$ & & $\asymm$ & & $2-\frac{2(N_c+2k)}{(k+1)N_f} + \frac{2r+1}{k+1}$\\
$\tilde P_r = \tilde Q A (\tilde A A)^r \tilde Q$ &  &  & $\asymm$& $2-\frac{2(N_c+2k)}{(k+1)N_f} + \frac{2r+1}{k+1}$
\end{tabular}
\ee
\end{center}
\bea
\nonumber
W = \Tr (a \tilde a )^{k+1} + \sum_{j=0}^k M_{k-j} q (\tilde a a)^j \tilde q + \sum_{r=0}^{k-1} P_{k-r-q} q (\tilde a a)^r \tilde a q + \tilde P_{k-r-q} \tilde q a (\tilde a a)^r \tilde q 
\eea
where the indices j runs from $0$ to $k$ and r runs from $0$ to $k-1$ and $\tilde N_c = (2k+1)N_f - 4k -N_c$.  As before, we can examine when a runaway can develop.  We find that the theory develops a runaway when $N_f < \frac{N_c + 2 k}{2k+1}$.  Unlike the previous case, this instability occurs after the dual gauge group has run negative.  Thus one expects that there is likely a richer set of confining dynamics in these theories as compared to the previous cases.


Using this duality, one can find the self-duality of the theory
\begin{center}
\be
\begin{tabular}{c|c|c}
&$SU(N)$&$SU(N+2-\delta)$\\
\hline
&&\\[-8pt]
$Q$ & $\fund$ & $\fund$  \\
$\tilde Q$ & $\overline \fund$ &  $\overline \fund$ \\
$A$ & $\asymm$ &  \\
$\tilde A$ & $\overline \asymm$  & 
\end{tabular}
\ee
\end{center}
\bea
\nonumber
W = \Tr (A \tilde A )^2 + (Q \tilde Q )^2 + \Tr (Q \tilde A  A \tilde Q)
\eea
It is dual to an identical gauge theory with gauge group $SU(N-2\delta)$ so that a self-duality is obtained for $\delta=0$.  For the self-dual gauge theory, the R charges of all fields is 1/2.

\subsection{$SO(N_c)$ with an antisymmetric}
\label{Sec: SO asym}

There is duality involving antisymmetrics and $SO$ gauge groups\cite{Leigh:1995qp} which is summarized as follows.  The electric theory is 
\begin{center}
\be
\begin{tabular}{c|c|c}
&$SO(N_c)$&$SU(N_f)$\\
\hline
&&\\[-8pt]
$Q$ & $\fund$ & $\fund$  \\
$A$ & $\asymm$ &   
\end{tabular}
\ee
\end{center}
\bea
\nonumber
W = \Tr A^{2k+2}
\eea
This theory is dual to 
\begin{center}
\be
\begin{tabular}{c|c|c}
&$SO(\tilde N_c)$&$SU(N_f)$\\
\hline
&&\\[-8pt]
$q$ & $\fund$ & $\overline \fund$ \\
$a$ & $\asymm$  & \\
$M_{j (\text{even)}} = Q a^j Q$ && $\symm$ \\
$M_{j (\text{odd)}} = Q a^j Q$ && $\asymm$
\end{tabular}
\ee
\end{center}
\bea
\nonumber
W = \Tr a^{2k+2} + \sum_{j=0}^{2k} M_{2k-j} q a^j \tilde q
\eea
where the index j can run from $0$ to $2k$ and $\tilde N_c = (2 k +1) N_f + 4 - N_c$.

Using Higgsing, one self-dual fixed point can be reached.  The UV is
\begin{center}
\be
\begin{tabular}{c|c|c}
&$SO(N)$&$SO(N-2-\delta)$\\
\hline
&&\\[-8pt]
$Q$ & $\fund$ & $\fund$  \\
$A$ & $\asymm$ &   
\end{tabular}
\ee
\end{center}
\bea
\nonumber
W = \Tr A^{4} + Q A^2 Q + Q^4
\eea
Applying the above duality and integrating out matter, this theory is dual to itself with gauge group $SO(N-2 \delta)$.

\subsection{$SO(N_c)$ with a symmetric}
\label{Sec: SO sym}

There is duality involving symmetric tensors and $SO$ gauge groups\cite{Intriligator:1995ff} which is summarized as follows.  The electric theory is 
\begin{center}
\be
\begin{tabular}{c|c|c}
&$SO(N_c)$&$SU(N_f)$\\
\hline
&&\\[-8pt]
$Q$ & $\fund$ & $\fund$  \\
$S$ & $\symm$ &   
\end{tabular}
\ee
\end{center}
\bea
\nonumber
W = \Tr S^{k+1}
\eea
This theory is dual to 
\begin{center}
\be
\begin{tabular}{c|c|c}
&$SO(\tilde N_c)$&$SU(N_f)$\\
\hline
&&\\[-8pt]
$q$ & $\fund$ & $\overline \fund$ \\
$s$ & $\symm$  & \\
$M_j = Q S^j Q$ && $\symm$
\end{tabular}
\ee
\end{center}
\bea
\nonumber
W = \Tr s^{k+1} + \sum_{j=0}^{k-1} M_{k-j-1} q x^j \tilde q
\eea
where the index j can run from $0$ to $k-1$ and $\tilde N_c = k N_f + 4 k - N_c$.  

Using Higgsing, two different self-dual fixed points can be reached.  The first is
\begin{center}
\be
\begin{tabular}{c|c|c}
&$SO(N)$&$SO(2N-8-\delta)$\\
\hline
&&\\[-8pt]
$Q$ & $\fund$ & $\fund$  \\
$S$ & $\symm$ &   
\end{tabular}
\ee
\end{center}
\bea
\nonumber
W = \Tr S^{3} + Q S Q
\eea
Applying the above duality and integrating out matter, this theory is dual to itself with gauge group $SO(N-\delta)$.

The second self-dual point is realized for
\begin{center}
\be
\begin{tabular}{c|c|c}
&$SO(N)$&$SO(N-6-\delta)$\\
\hline
&&\\[-8pt]
$Q$ & $\fund$ & $\fund$  \\
$S$ & $\symm$ &   
\end{tabular}
\ee
\end{center}
\bea
\nonumber
W = \Tr S^{4} + Q S^2 Q + Q^4
\eea
Applying the above duality and integrating out matter, this theory is dual itself with gauge group $SO(N-2\delta)$.

\subsection{$Sp(N_c)$ with an antisymmetric}
\label{Sec: Sp asym}

There is duality involving antisymmetrics and $Sp$ gauge groups\cite{Intriligator:1995ff} which is summarized as follows.  The electric theory is 
\begin{center}
\be
\begin{tabular}{c|c|c}
&$Sp(2 N_c)$&$SU(2 N_f)$\\
\hline
&&\\[-8pt]
$Q$ & $\fund$ & $\fund$  \\
$A$ & $\asymm$ &   
\end{tabular}
\ee
\end{center}
\bea
\nonumber
W = \Tr A^{k+1}
\eea
This theory is dual to 
\begin{center}
\be
\begin{tabular}{c|c|c}
&$Sp(2 \tilde N_c)$&$SU(2 N_f)$\\
\hline
&&\\[-8pt]
$q$ & $\fund$ & $\overline \fund$ \\
$a$ & $\asymm$  & \\
$M_j = Q A^j Q$ && $\asymm$
\end{tabular}
\ee
\end{center}
\bea
\nonumber
W = \Tr a^{k+1} + \sum_{j=0}^{k-1} M_{k-j-1} q a^j \tilde q
\eea
where the index j can run from $0$ to $k-1$ and $\tilde N_c = k N_f - 2 k - N_c$.  

Using Higgsing, two different self-dual fixed points can be reached.  The first is
\begin{center}
\be
\begin{tabular}{c|c|c}
&$Sp(2 N)$&$Sp(4 N +8 - 2 \delta)$\\
\hline
&&\\[-8pt]
$Q$ & $\fund$ & $\fund$  \\
$A$ & $\asymm$ &   
\end{tabular}
\ee
\end{center}
\bea
\nonumber
W = \Tr A^{3} + Q A Q
\eea
Applying the above duality and integrating out matter, this theory is dual to itself with gauge group $Sp(2 N - 2 \delta)$.

The second self-dual point is realized for
\begin{center}
\be
\begin{tabular}{c|c|c}
&$Sp(2 N)$&$Sp(2 N + 6 - 2 \delta)$\\
\hline
&&\\[-8pt]
$Q$ & $\fund$ & $\fund$  \\
$A$ & $\asymm$ &   
\end{tabular}
\ee
\end{center}
\bea
\nonumber
W = \Tr A^{4} + Q A^2 Q + Q^4
\eea
Applying the above duality and integrating out matter, this theory is dual to itself with gauge group $Sp(2 N - 4 \delta)$.

\subsection{$Sp(N_c)$ with a symmetric}
\label{Sec: Sp sym}

There is duality involving symmetrics and $Sp$ gauge groups\cite{Leigh:1995qp} which is summarized as follows.  The electric theory is 
\begin{center}
\be
\begin{tabular}{c|c|c}
&$Sp(2 N_c)$&$SU(2 N_f)$\\
\hline
&&\\[-8pt]
$Q$ & $\fund$ & $\fund$  \\
$S$ & $\symm$ &   
\end{tabular}
\ee
\end{center}
\bea
\nonumber
W = \Tr S^{2k+2}
\eea
This theory is dual to 
\begin{center}
\be
\begin{tabular}{c|c|c}
&$Sp(2 \tilde N_c)$&$SU(2 N_f)$\\
\hline
&&\\[-8pt]
$q$ & $\fund$ & $\overline \fund$ \\
$s$ & $\symm$  & \\
$M_{j (\text{even)}} = Q s^j Q$ && $\asymm$ \\
$M_{j (\text{odd)}} = Q s^j Q$ && $\symm$
\end{tabular}
\ee
\end{center}
\bea
\nonumber
W = \Tr s^{2k+2} + \sum_{j=0}^{2k} M_{2k-j} q s^j \tilde q
\eea
where the index j can run from $0$ to $2k$ and $\tilde N_c = (2k+1) N_f - 2 - N_c$.  

Using Higgsing, a self-dual fixed point can be reached.  It is
\begin{center}
\be
\begin{tabular}{c|c|c}
&$Sp(2 N)$&$Sp(2 N + 2 - 2 \delta)$\\
\hline
&&\\[-8pt]
$Q$ & $\fund$ & $\fund$  \\
$S$ & $\symm$ &   
\end{tabular}
\ee
\end{center}
\bea
\nonumber
W = \Tr S^{4} + Q S^2 Q + Q^4
\eea
Applying the above duality and integrating out matter, this theory is dual to itself with gauge group $Sp(2N-4\delta)$.

\section{A vector-like cascade with meta-stable SUSY breaking}
\label{App: vector}

We provide another duality cascade which shows that the problem of incalculable singlets can be cured in the context of a vector-like cascade.  
The starting gauge theory is 
\begin{center}
\be
\begin{tabular}{c|cc|}
&$SU(N)$&$SO(N+2-\delta)$\\
\hline
&&\\[-8pt]
$Q$ & $\fund$ & $\fund$  \\
$\tilde Q$ & $\overline \fund$ &  $\fund$ \\
$A$ & $\asymm$ &  \\
$\tilde A$ & $\overline \asymm$  & 
\end{tabular}
\ee
\end{center}
\bea
\nonumber
W = \Tr (A \tilde A )^2 + (Q \tilde Q )^2 + \Tr (Q \tilde A  A \tilde Q)
\eea
After two applications of duality, the theory goes back to the original theory with $N' = N - 2 \delta$.  Assume that $2 \delta > N > \frac{3 \delta -2}{2}$ so that the $SU(N)$ theory ends the cascade through supersymmetry breaking.  The dual gauge theory is 
\begin{center}
\be
\begin{tabular}{c|cc|}
&$SU(2N+2-3\delta)$&$SO(N+2-\delta)$\\
\hline
&&\\[-8pt]
$q$ & $\fund$ & $\fund$  \\
$\tilde q$ & $\overline \fund$ &  $\fund$ \\
$a$ & $\asymm$ &  \\
$\tilde a$ & $\overline \asymm$  & \\
$M_0$ &   & $\fund^2$ \\
$M_1$ &   & $\fund^2$ \\
$P$ &   & $\asymm$ \\
$\tilde P$ &   & $\asymm$
\end{tabular}
\ee
\end{center}
\bea
\nonumber
W = \Tr (a \tilde a )^2 + M_0^2 + M_1 + M_1 q \tilde q + M_0 q \tilde a a \tilde q + P q \tilde a q + \tilde P \tilde q a \tilde q
\eea
As before, supersymmetry is broken through the rank condition by the F term of $M_1$.  $M_0$ has a tree level mass while $M_1$ and $q$, $\tilde q$ all obtain a positive mass.  There is no singlet meson in $P$ so that they all receive positive two-loop masses from gauge mediation.  Thus we find that this vector-like theory has a meta-stable SUSY breaking vacuum without the singlet problem.

\section{Gaugino condensation and ADS superpotentials}
\label{App: example}

Throughout this paper, we take symmetry breaking vevs and use gaugino condensation to understand the behavior of the theory.  In this appendix we show how the ADS superpotential can be derived using the same techniques.  We start with SQCD with $N_f < N_c - 1$.  For simplicity we explore the D flat direction
\bea
\langle Q \rangle =  q \begin{pmatrix}
  \mathds{1}_{N_f \times N_f}  \\
   0_{N_c - N_f \times N_f}
 \end{pmatrix}  &\qquad&
\langle \tilde Q \rangle =  q \begin{pmatrix}
  \mathds{1}_{N_f \times N_f}  \\
   0_{N_c - N_f \times N_f}
 \end{pmatrix}  
\eea
The remaining gauge group undergoes gaugino condensation.  After using scale matching we find the superpotential
\bea
W \sim ( \Lambda^{3(N_c - N_f)}_{N_c - N_f})^\frac{1}{N_c - N_f} \sim ( \frac{\Lambda^{3 N_c - N_f}_{N_c}}{q^{2 N_f}})^\frac{1}{N_c - N_f}
\eea
Using symmetry rotations, the denominator can be expressed as the familiar determinant term.  We have rederived the ADS superpotential using gaugino condensation.  Notice that this derivation requires that $N_c - N_f \ge 2$.  For $N_c = N_f +1$ instanton calculations are required to show the existence of the superpotential.

\end{document}